\documentclass{elsart3p}

 \usepackage{graphics}
 \usepackage{graphicx}

\usepackage{amssymb}

\begin{document}

\begin{frontmatter}



\title{Point-contact spectroscopy of the nickel borocarbide
superconductors $R$Ni$_{2}$B$_{2}$C ($R$=Y, Dy, Ho, Er, Tm, Lu)}


\author[a]{Yu. G. Naidyuk},
\author[a]{D. L. Bashlakov},
\author[a]{N. L. Bobrov},
\author[a]{V. N. Chernobay},
\author[a]{O. E. Kvitnitskaya},
\author[a]{I. K. Yanson},
\author[b]{G. Behr},
\author[b]{S.-L. Drechsler},
\author[b]{G. Fuchs},
\author[b]{D. Souptel},
\author[c]{D. G. Naugle},
\author[c]{K. D. D. Rathnayaka},
\author[c]{J. H. Ross Jr.}

\address[a]{B. Verkin Institute for Low Temperature Physics and
Engineering, NAS of Ukraine,  47 Lenin Ave., 61103, Kharkiv,
Ukraine}
\address[b]{Leibniz-Institut f\"{u}r Festk\"{o}rper- und
Werkstoffforschung Dresden e.V., Postfach 270116, D-01171 Dresden,
Germany}
\address[c]{Department of Physics, Texas A\&M University, College Station TX 77843-4242,
USA}

\begin{abstract}
An overview of the recent efforts in point-contact (PC)
spectroscopy of the nickel borocarbide superconductors
$R$Ni$_{2}$B$_{2}$C in the normal and superconducting (SC) state
is given. The  results of measurements of the PC
electron-boson(phonon) interaction spectral function are
presented. Phonon maxima and crystalline-electric-field (CEF)
excitations are observed in the PC spectra of compounds with
$R$=Dy, Ho, Er and Tm, while for $R$=Y a dominant phonon maximum
around 12\,meV is characteristic. Additionally, non-phonon and
non-CEF maxima are observed near 3\,meV in $R$=Ho and near 6\,meV
in $R$=Dy. Directional PC study of the SC gap gives evidence for
the multi-band nature of superconductivity in $R$=Y, Lu. At low
temperature the SC gap in $R$=Ho exhibits a standard single-band
BCS-like dependence, which vanishes above
$T_{c}^{*}\simeq5.6$\,K$<T_{c}\simeq 8.5\,$K, where a specific
magnetic ordering starts to play a role. For $R$=Tm ($T_{c}\simeq
10.5\,$K) a decrease of the SC gap is observed below 5\,K.


\end{abstract}

\end{frontmatter}

The $R$Ni$_{2}$B$_{2}$C ($R$ is mainly rare earth, Y  or Sc)
family of ternary superconductors discovered in 1994 has become
attractive to study fundamental questions in superconductivity
\cite{Muller}. The interest is because the critical temperature in
$R$Ni$_{2}$B$_{2}$C is relatively high, their superconducting (SC)
properties exhibit often unconventional behavior and depending on
the rare earth element magnetism and superconductivity co-exist in
these materials in a wide range of temperatures. Last but not
least there is continious progress in synthesis of high purity
single crystal samples. However in spite of dozens of publications
some fundamental issues as to the nature of Cooper pairing and
attractive interaction are still under debate for the
borocarbides.

By point-contact (PC) research both the SC gap and the PC
electron-phonon(boson) interaction (EP(B)I) function
$\alpha^2_{\rm PC}F(\omega)$ can be determined from the first and
second derivatives of the $I(V)$ characteristic of PC's
\cite{Naid}. The second derivative of $I(V)$ contains
straightforward information as to the PC EB(P)I spectral function
$\alpha^{2}F(\omega)$ \cite{Naid}. Knowing $\alpha^{2}F(\omega)$
one can elucidate the nature of the attractive interaction and
characteristic bosons. Unique is that with the same PCs driven to
the superconducting state the SC gap can be measured. Thus the PC
spectroscopy is a powerful method to study both the EP(B)I spectra
and the SC gap behavior.

\begin{figure}
\begin{center}
\includegraphics[width=7cm,angle=0]{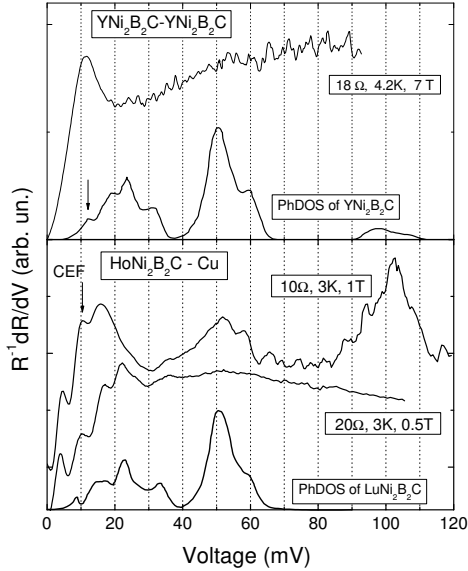}
\end{center}
\caption{PC spectra of $R$Ni$_{2}$B$_{2}$C ($R$=Y, Ho)  in
comparison  with the phonon DOS for YNi$_{2}$B$_{2}$C and
LuNi$_{2}$B$_{2}$C \cite{Gompf}. The superconductivity which
causes huge features in the low bias region is suppressed by a
magnetic field. For the Ho compounds two kind of spectra are
shown.} \label{f1}
\end{figure}

\begin{figure}
\begin{center}
\includegraphics[width=7cm,angle=0]{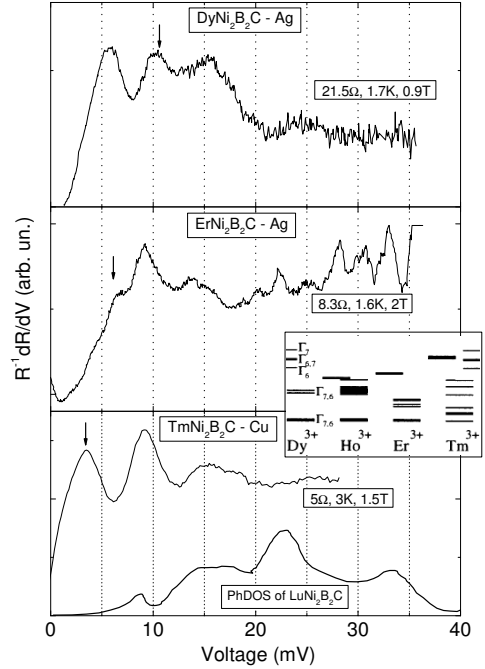}
\end{center}
\caption{Low energy part of the PC spectra of $R$Ni$_{2}$B$_{2}$C
($R$=Dy, Er, Tm)) in comparison  with the phonon DOS for
LuNi$_{2}$B$_{2}$C \cite{Gompf}. The superconductivity is
suppressed by a magnetic field. The arrows show approximate
position of the first exited CEF level for each compound according
to the CEF scheme from \cite{Gasser} shown in the inset.}
\label{f2}
\end{figure}

\begin{figure}
\begin{center}
\includegraphics[width=8cm,angle=0]{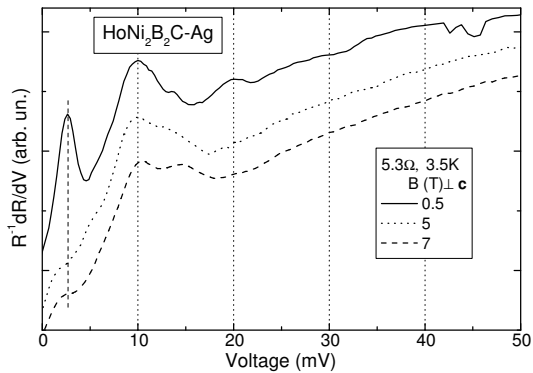}
\end{center}
\caption{PC spectra of a HoNi$_{2}$B$_{2}$C--Ag contact which
demonstrate pronounced 3 and 10\,mV-maxima and their modification in
a magnetic field. } \label{f3}
\end{figure}

At first we start with the discussion of PC EP(B)I spectra in
$R$Ni$_{2}$B$_{2}$C presented in Figs.\,\ref{f1} and \ref{f2}. The
most detailed spectra are measured for HoNi$_{2}$B$_{2}$C. Here,
most of the maxima correspond to those in the phonon DOS of the
isostructural compound LuNi$_2$B$_2$C \cite{Gompf}. The upper
Ho-spectrum displays also expressed high energy maxima around 50
and 100\,mV which were not so clearly resolved so far in the PC
spectra of other nickel borocarbides. The low energy maximum
around 3\,mV in $R$=Ho and 6\,mV in $R$=Dy (Fig.\,\ref{f2}), not
present in the phonon DOS, have non phonon origins. The 3-mV
maximum in $R$=Ho can be suppressed by a magnetic field
(Fig.\,\ref{f3}) pointing to its "magnetic" origin as discussed in
\cite{NaidSCES,NaidM2S}. The 6-mV peak in $R$=Dy vanishes with
increasing $T$ above 15\,K, has probably a similar "magnetic"
origin. The first maximum in $R$=Tm and $R$=Er corresponds in
position to the first exited crystal-electric-field (CEF) level
\cite{Gasser}. CEF contributes apparently also to the 10-mV peak
in $R$=Ho and Dy compounds. In the former case this is seen from
the modification of the 10-mV peak in a magnetic field
(Fig.\,\ref{f3}). Most of the PC spectra in $R$=Ho (not shown)
demonstrate a prominent 10-mV peak while other phonon maxima are
completely smeared. This points to the importance of the CEF
excitations in the charge transport as well as in the SC
properties of HoNi$_2$B$_2$C and other $R$=Dy, Er and Tm
borocarbides. Thus the contribution of the 10-mV peak in the
EP(B)I constant $\lambda_{PC}$ for $R$=Ho is evaluated as 20-30\%,
while the contribution to $\lambda_{PC}$ of the high frequency
modes at 50 and 100\,meV amounts up to 10\% for each maximum.

The spectra of the nonmagnetic YNi$_2$B$_2$C show a dominant
maximum at about 12\,mV and a broad shallow maximum or a kink
positioned close to 50\,mV (Fig.\,\ref{f1}). These features have a
counterpart in the phonon DOS of YNi$_2$B$_2$C. To summarize: a
clear coupling to the low energy modes has been shown in the PC
spectra of the title compounds. In $R$=Dy, Ho, Er and Tm compounds
the CEF excitations contribute to the EP(B)I function. For $R$=Ho
the contribution of two high energy modes are also notable. The
50-mV mode can be also resolved for other compounds, but it looks
very smeared as shown in Fig.\,\ref{f1} for $R$=Y. In this
context, we note that the electronic mean free path is shortened
with increase of the bias voltage due to the EP(B)I. In this case
if the EP(B)I is strong enough, it results in a violation of the
ballistic condition (with voltage increase) necessary for energy
resolved spectroscopy by PCs. On the other hand the high energy
modes involve vibration of light B and C ions. As it is known, the
physical properties of nickel borocarbides, including the
superconductivity, are sensitive to small variations in the
crystal stoichiometry. Disorder in the B and C position is
especially difficult to avoid and control. Turning to the PC
spectra we mention that the pronounced high energy maxima for the
Ho compound were measured for a sample with improved RRR and
superconductivity after a special annealing treatment.

\begin{figure}
\begin{center}
\includegraphics[width=6cm,angle=0]{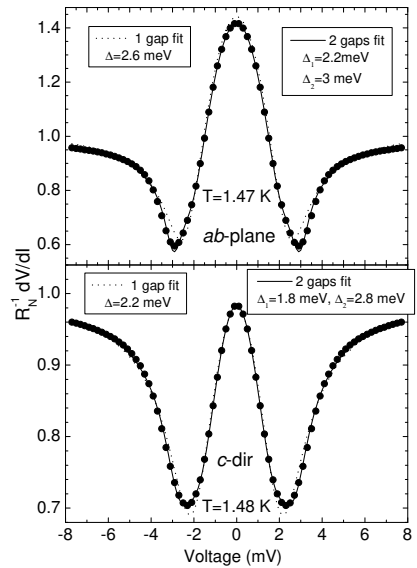}
\end{center}
\caption{Fit of $dV/dI$ (solid circles) of LuNi$_{2}$B$_{2}$C--Ag
point contacts established in the ab-plane and along the c-axis in
the one-gap (dotted curves) and two-gap (solid curves)
approximation \cite{Bobrov}. $R_N$=22.5\,$\Omega$ (upper curve)
and 45\,$\Omega$ (bottom curve).}
 \label{f4}
\end{figure}

The SC gap manifests itself in the $dV/dI$ characteristic of a
N-c-S contact as pronounced minima symmetrically placed at
$V\simeq\pm\Delta$ for $T\ll T_{c}$. In the case of a
two-gap(band) superconductor like MgB$_2$ (and firstly proposed
for borocarbides already in 1998 \cite{Shulga}) a number of minima
are seen, each pair corresponding to a definite gap (see e.\,g.
\cite{Naid}). Up to now all measured $dV/dI$ curves for SC
borocarbides exhibit only one pair of minima as in the case of a
single gap. However as it is shown in \cite{Bobrov}, a two gap
approach allows better to fit the experimental curves for
LuNi$_2$B$_2$C (Fig.\,\ref{f4}). The reason why only one gap
minimum structure is visible is that the gap values differ only by
a factor of 1.5 while in MgB$_2$ this difference is about 3 times.

\begin{figure}
\begin{center}
\includegraphics[width=6cm,angle=0]{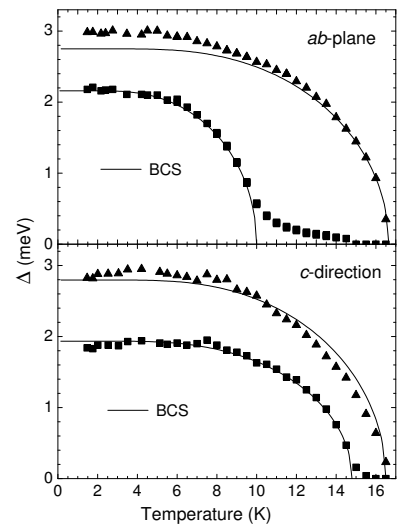}
\end{center}
\caption{$\Delta(T)$ behavior in LuNi$_{2}$B$_{2}$C obtained in
\cite{Bobrov} by fitting of the $dV/dI$ curves in the two-gap
approximation. The solid curves show the BCS-like gap behavior.}
 \label{f5}
\end{figure}

It is interesting that in the BCS extrapolation the critical
temperature, $T_c$, corresponding to the small gap in
LuNi$_2$B$_2$C is 10\,K in the ab plane and 14.5\,K in the c
direction (see Fig.\,\ref{f5}). Note the unusual temperature
dependence of this gap suggesting extremely weak interband
coupling (compare with \cite{Walte}). For the large gap $T_c$
coincides with the bulk $T_c$=16.8\,K and the absolute values are
about 3\,meV in both orientations. In the c direction the
contributions to the conductivity from the small and the large
gaps remain practically identical up to 10–-11\,K. In the ab plane
the contribution from the small gap is smaller and decreases
rapidly with increasing $T$.

The single gap approach used in \cite{NaidM2S} gives an
anisotropic gap in YNi$_2$B$_2$C with a gap of about 1.5\,meV
along the a-axis, a larger 2.3\,meV gap along the c-axis and
maximal 2.5\,meV gap in the [110] direction. The last two values
are close to the corresponding values in LuNi$_2$B$_2$C for the
single gap fit shown in Fig.\,\ref{f4}. Thus the measurements on
Lu and Y compounds are compatible with a multi(two) band scenario
of the SC state in these both nonmagnetic borocarbides.

\begin{figure}
\begin{center}
\includegraphics[width=8cm,angle=0]{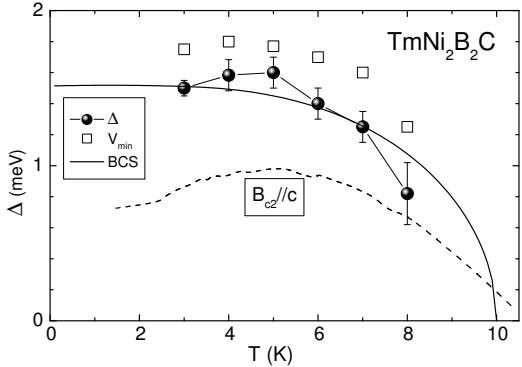}
\end{center}
\caption{$\Delta(T)$ behavior in TmNi$_{2}$B$_{2}$C established by
a BTK fit of $dV/dI$ curves. The squares show the position of the
minimum in $dV/dI$. The dashed curve shows the upper critical
field, $B_{c2}$, in TmNi$_{2}$B$_{2}$C for $B$ along the c axis. }
\label{f6}
\end{figure}

Turning to the SC gap measurement in the Ho compound, no obvious
gap anisotropy was observed \cite{NaidM2S}. The intriguing feature
here is that the gap vanishes at $T_{c}^{*}\simeq$5.6\,K, which is
close to the N$\acute{e}$el temperature $T_{N}\simeq$5.3\,K, but
well below $T_{c}\simeq$8.5\,K of the bulk crystal. This suggest
that the superconductivity in the commensurate antiferromagnetic
phase of the Ho compound survives at a special nearly isotropic
Fermi Surface Sheet. Between $T_{c}^{*}$ and $T_{c}\simeq$8.5\,K
the SC signal in $dV/dI$ is drastically suppressed, giving no
possibility to observe a finite SC gap by fitting the $dV/dI$
data. This shows that the SC state in HoNi$_{2}$B$_{2}$C between
$T_{c}^{*}$ and $T_{c}$, i.\,e. in the region of peculiar
incommensurate magnetic order, is unusual.

An interesting nonmonotonic behavior characterizes the $\Delta(T)$
in TmNi$_{2}$B$_{2}$C (Fig.\,\ref{f6}) at the  maximum around 5\,K
where the gap decreases again with lowering of $T$. This is
similar to the behavior of the upper critical field along the
c-axis. Apparently, antiferomagnetic fluctuations due to the
approaching of a N$\acute{e}$el phase $T_N\simeq$1.5\,K are
responsible for the gap decrease.

\begin{figure}
\begin{center}
\includegraphics[width=8cm,angle=0]{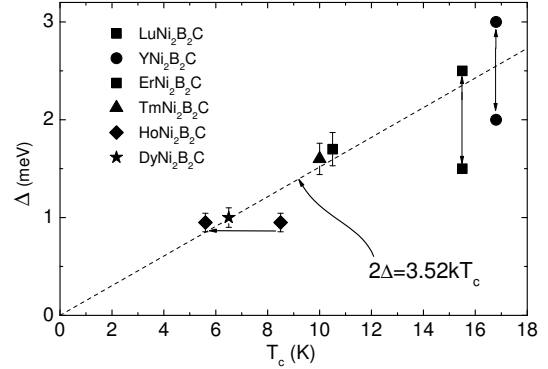}
\end{center}
\caption{The gap $\Delta$ (symbols) as a function of the critical
temperature $T_c$ in $R$Ni$_{2}$B$_{2}$C established by a PC
study. For $R$=Y the extrema of the anisotropic gap are presented
while for $R$=Lu the small and large gaps are shown. For $R$=Ho
$T_c$ is shifted to $T_c^*$=5.6\,K (see text for explanation). The
BCS ratio is shown by the dashed straight line.} \label{f7}
\end{figure}

Figure\,\ref{f7} summarizes the measurement of the SC gap by PCs
in the title compounds. In general, the gaps are placed close to
the BCS value $2\Delta/k_B T_c\simeq 3.52$ taking into account the
anisotropy or multiband behavior in the Y and Lu compounds and the
specific SC state in the Ho borocarbide.

Support by SFB 463 (Germany),  US CRDF (No.UP1-2566-KH-03)
grants and NAS of Ukraine are acknowledged. Work at TAMU was partially
supported by the Robert A. Welch Foundation, Houston, TX (A-0515, A-1526).

\end{document}